\documentclass[10pt]{article}

\begin{document}

\title{Wave-like Solutions for Bianchi type-I cosmologies in $5D$}
\author{J. Ponce de Leon\thanks{E-Mail:
jpdel@ltp.upr.clu.edu, jpdel1@hotmail.com}  \\Laboratory of Theoretical Physics, 
Department of Physics\\ 
University of Puerto Rico, P.O. Box 23343,  
San Juan,\\ PR 00931, USA}
\date{July, 2008}

\maketitle
\begin{abstract}

We derive exact solutions to the vacuum Einstein field equations in $5D$, under the assumption that (i) the line element in $5D$ possesses self-similar symmetry, in the classical understanding of Sedov, Taub and Zeldovich,  and that (ii) the metric tensor is diagonal and independent of the coordinates for ordinary $3D$ space. These assumptions lead to three different types of self-similarity in $5D$: homothetic, conformal and ``wave-like".
In this work we present the most general wave-like solutions to the $5D$ field equations. Using the standard technique based on Campbell's theorem,  they generate a large   number of anisotropic cosmological models of Bianchi type-I, which can be applied to our universe after the big-bang, when anisotropies could have played an important role. We present a complete review of all possible cases of self-similar anisotropic cosmologies in $5D$. Our analysis extends a number of previous studies on wave-like solutions in $5D$ with spatial spherical symmetry.

\end{abstract}

\medskip



\newpage
\section{Introduction}
Nowadays, there are a number of theories suggesting that the universe may have more than  four dimensions. Extra dimensions arise naturally in supergravity $(11D)$ and superstring theories $(10D)$, which seek the unification of gravity with the interactions of particle physics, and are expected to become important at very high energies, e.g., near the horizon of black holes \cite{Davidson Owen} and during the  evolution of the early universe \cite{evol of early universe}. 

Therefore, it is worthwhile to explore cosmological models in presence of extra dimensions. In this regard, a powerful theoretical tool is provided by Campbell's theorem \cite{Campbell}, \cite{Seahra}, which serves as a ladder to go between manifolds whose dimensionality differs by one. This theorem, which is valid in any number of dimensions, implies that every solution of the $4D$ Einstein equations with arbitrary energy-momentum tensor can be embedded, at least locally, in a solution of  the five-dimensional \underline{vacuum} Einstein field equations. In this work we will derive exact solutions to the $5D$ field equations which embed a large family of anisotropic cosmological models of Bianchi type-I that may be applicable to the early universe. 

In conventional $4D$ general relativity, in order to solve the field  equations one usually assumes a form for the matter content, i.e., the energy-momentum tensor,  and imposes certain symmetries on the spacetime. For example, if we  assume empty space and spatial spherical symmetry we obtain the Schwarzschild solution;  if we assume that the matter satisfies a barotropic equation of state, and that the spacetime is homogeneous and isotropic, then we obtain the standard Friedmann-Robertson-Walker (FLW) cosmological models.

An important consequence of Campbell's theorem is, in particular,  that  for the study of cosmological models embedded in $5D$ we do not need a five-dimensional energy-momentum tensor. Thus, instead of seeking an embedding of a $4D$ spacetime with a specified physical energy-momentum tensor, 
	the procedure in $5D$ is as follows: first  one has to find a solution to  the fifteen Einstein field equations in vacuum,  then the properties of the $4D$ \underline{effective} matter source  for the $5D$ solutions are deduced after choosing an embedding. The standard technique\footnote{This standard technique is used in braneworld \cite{algebraically} as well as in space-time-matter (or induced matter) theory \cite{general}. Both theories employ a $5D$ Kaluza-Klein type of metric, $dS^2 = \gamma_{AB}dx^{A}dx^{B} = g_{\mu\nu}(x^{\rho}, \psi)dx^{\mu}dx^{\nu} + \epsilon \Phi(x^{\rho}, \psi)d\psi^2$, where the extra dimension $\psi$ is not assumed to be compactified as in the original account. Consequently, both theories are mathematically equivalent in the sense that they lead to the same {\it effective} energy-momentum tensor in $4D$, although they have different  physical interpretation \cite{physical motivation}.} consists in isolating the $4D$ part of the relevant $5D$ quantities, using them to construct the $4D$ Einstein tensor $G_{\alpha \beta}$ $(\alpha, \beta = 0, 1, 2, 3)$ and utilizing the field equations of general relativity $G_{\alpha \beta } = 8 \pi T_{\alpha \beta}$ (we use $c = G = 1$) to identify the effective energy-momentum tensor $T_{\alpha \beta}$.

At this point, the  natural question is how can we construct  solutions to the $5D$ vacuum Einstein equations that lead to models with `reasonable' physical properties in $4D$. Clearly, the best approach to accomplish this is to impose spacetime symmetries,  on the $5D$ metric, that are characteristic of the $4D$ source  that we want to embed in $5D$. This is illustrated by a number of $5D$ solutions, e.g., the  Kramer-Gross-Perry-Davidson-Owen solutions \cite{Davidson Owen}, \cite{Kramer} which embed the Schwarzschild solution of general relativity; the `standard' $5D$ cosmologies \cite{standard} that reduce to the usual FRW cosmologies with flat space sections, on every hypersurface defined by fixing the fifth coordinate (which we denote as $\psi$). This approach allows not only to recover known solutions of $4D$ general relativity, but also generates  new ones that may shed some light on the effects of a putative extra dimension on the physics in $4D$.  

Following this {\it modus operandi}, it is important to investigate  five-dimensional cosmological models whose metric tensor is diagonal and independent of the coordinates for ordinary $3D$ space. This is  because they reduce to  homogeneous cosmological models with flat spatial sections on every hypersurface $\Sigma_{\psi}: \psi = \psi_{0}$ and encompass anisotropic cosmological models of Bianchi type-I as well as flat FRW models.
In a recent work \cite{Anis Cosm I} we started a systematic investigation of such models under the assumption that the line element\footnote{Notation: here the coordinates are assigned as usual, $x^0 = t$ for time; $x^{1, 2, 3} = x, y, z$ for space; $x^{4} = \psi$ for the extra coordinate, and $\epsilon = \pm 1$ depending on whether the extra dimension is spacelike or timelike.} in $5D$ 

\begin{equation}
\label{General metric in 5D }
dS^2 = e^{\nu(t, \psi)}dt^2 - e^{\lambda(t, \psi)}dx^2 - e^{\mu(t, \psi)}dy^2 - e^{\sigma(t, \psi)}dz^2 + \epsilon  e^{\omega(t, \psi)}d\psi^2,
\end{equation}
admits  self-similar symmetry, in the sense that all the dimensionless quantities are assumed to be functions of a single variable $\xi$ \cite{Sedov}, which is some combination of the coordinates $x^{0} = t$ and $x^{4} = \psi$. In this way the field equations become a system of ordinary, instead of partial, differential equations. From a physical point of view,  the assumption of self-similarity  is motivated by a number of studies suggesting  that many homogeneous and inhomogeneous cosmological models can be approximated by self-similar homothetic models in the asymptotic regimes \cite{Coley}, i.e., near the initial cosmological singularity and at late times.  
In \cite{Anis Cosm I} we showed that there are three possible choices for the similarity variable which lead to solutions with  different physical and mathematical properties. These are (i) $\xi = t/\psi$; (ii) $\xi = e^{q t}/e^{q \psi}$, and (iii) $\xi = {\cal{E}} t + k \psi$, where $q, {\cal{E}}$ and $k$ are some constants with the appropriate units. We found the most general solutions to the $5D$ vacuum field equations $R_{A B} = 0$ $(A, B = 0, 1, 2, 3, 4)$ corresponding to the first two choices of $\xi$ and showed that they admit  homothetic and conformal symmetry in $5D$, respectively.  

However, in \cite{Anis Cosm I} we did not discuss the third case where the metric functions have a dependence of time and the extra coordinate like in traveling waves or pulses propagating along the fifth dimension. In this work we conclude our analysis of self-similar anisotropic cosmologies in $5D$ by giving a detailed discussion of the field equations and their solutions for the case where  the similarity variable is $\xi = {\cal{E}} t + k \psi$. In short, we will refer to them as ``wave-like" solutions. This work extends a number of previous studies of wave-like solutions in $5D$ with spatial spherical symmetry \cite{extension}.

The paper is organized as follows. In section $2$ we present the general integration of the field equations. We will see that the solutions are expressed in terms of one arbitrary function and three dimensionless parameters. In sections $3$ we study some particular solutions which are generated by geometrical considerations. In section $4$ we close the system of equations by making certain assumptions on the ``equations of state" for the   $4D$ effective matter.  Finally, in section $5$ we present a summary and a complete analysis of all possible cases of self-similar anisotropic cosmologies in $5D$.

\section{Integrating the field equations}

Let us consider the five-dimensional, self-similar, line element
\begin{equation}
dS^2 = e^{\nu(\xi)}dt^2 -  e^{\lambda(\xi)}dx^2 - e^{\mu(\xi)}dy^2 - e^{\sigma(\xi)}dz^2 + \epsilon e^{\omega(\xi)}d\psi^2,
\end{equation}
with
\begin{equation}
\label{similarity variable}
\xi = {\cal{E}} t + k\psi,
\end{equation}
where ${\cal{E}}$ and $k$ are some constants. The five metric functions 
$\nu(\xi)$, $\lambda(\xi)$, $\mu(\xi)$, $\sigma(\xi)$, $\omega(\xi)$, as well as the signature of the extra coordinate,  have to be determined from the field equations $R_{AB} = 0$. The $5D$ Ricci tensor has six nonvanishing components, viz., $R_{00}, R_{11}, R_{22}, R_{33}, R_{44}, R_{04}$. However, not all of them are independent. We will see bellow that they reduce to two independent equations for three unknown functions.   

Setting $R_{11}, R_{22}, R_{33}$ to zero we obtain

\begin{equation}
\label{R11}
\lambda_{\xi}\left[\left(\epsilon {\cal{E}}^2 e^{\omega} + k^2 e^{\nu}\right)\left(\lambda_{\xi} + \frac{2 \lambda_{\xi\xi}}{\lambda_{\xi}} + \mu_{\xi} + \sigma_{\xi} \right) + \left(\epsilon {\cal{E}}^2 e^{\omega} - k^2e^{\nu}\right)\left(\omega_{\xi} - \nu_{\xi}\right)\right] = 0.
\end{equation}

\bigskip

\begin{equation}
\label{R22}
\mu_{\xi}\left[\left(\epsilon {\cal{E}}^2 e^{\omega} + k^2 e^{\nu}\right)\left(\mu_{\xi} + \frac{2 \mu_{\xi\xi}}{\mu_{\xi}} + \lambda_{\xi} + \sigma_{\xi}\right) + \left(\epsilon {\cal{E}}^2 e^{\omega} - k^2 e^{\nu}\right)\left(\omega_{\xi} - \nu_{\xi}\right)\right] = 0.
\end{equation}

\bigskip

\begin{equation}
\label{R33}
\sigma_{\xi}\left[\left(\epsilon {\cal{E}}^2 e^{\omega} + k^2 e^{\nu}\right)\left(\sigma_{\xi} + \frac{2 \sigma_{\xi\xi}}{\sigma_{\xi}} + \lambda_{\xi} + \mu_{\xi}\right) + \left(\epsilon {\cal{E}}^2 e^{\omega} - k^2e^{\nu}\right)\left(\omega_{\xi} - \nu_{\xi}\right)\right] = 0.
\end{equation} 

\bigskip
These equations require
\begin{equation}
\label{compatibility condition}
\frac{\lambda_{\xi\xi}}{\lambda_{\xi}} = \frac{\mu_{\xi \xi}}{\mu_{\xi}} = \frac{\sigma_{\xi \xi}}{\sigma_{\xi}}.
\end{equation}
Therefore, without loss of generality one can set
\begin{equation}
\label{introduction of f}
e^{\lambda} = A f^{2 \alpha}(\xi), \;\;\;e^{\mu} = B f^{2 \beta}(\xi), \;\;\;e^{\sigma} = C f^{2 \gamma}(\xi),
\end{equation}
where $A, B, C$ are constants;  $f$ is some function of the variable $\xi = ({\cal{E}} t + k \psi)$;  and  $\alpha$, $\beta$ and $\gamma$ are arbitrary parameters.  As a consequence, $R_{11} = 0$, $R_{22} = 0$ and $R_{33} = 0$ reduce to
\begin{equation}
\label{R11 = R22= R33}
\left(\epsilon {\cal{E}}^ 2 e^{\omega} + k^2 e^{\nu}\right)\left[\frac{f_{\xi \xi}}{f_{\xi}}+ \frac{f_{\xi}}{f}\left(\alpha + \beta + \gamma - 1\right)\right] + \frac{1}{2}\left(\epsilon {\cal{E}}^2 e^{\omega} - k^2 e^{\nu}\right)\left(\omega_{\xi} - \nu_{\xi}\right)= 0.
\end{equation}
On the other hand $R_{04} = 0$ yields
\begin{equation}
\label{R04 in terms of f}
(\alpha + \beta + \gamma)\left(\frac{2 f_{\xi \xi}}{f_{\xi}} - \nu_{\xi} -  \omega_{\xi}\right) + 2(\alpha^2 + \beta^2 + \gamma^2 - \alpha - \beta - \gamma)\left(\frac{f_{\xi}}{f}\right) = 0,
\end{equation}
from which we get 
\begin{equation}
\label{integration of R04}
e^{(\nu + \omega)/2} = E f^{(b/a)}f_{\xi},
\end{equation}
where $E$ is a constant of integration, and\footnote{Our equations (\ref{R11 = R22= R33})-(\ref{integration of R04}) are the counterparts of equations $(22)-(24)$ in \cite{Anis Cosm I} . Although they look alike, these are distinct differential equations.}
\begin{equation}
\label{def. of a and b}
a \equiv (\alpha + \beta + \gamma), \;\;\;b \equiv (\alpha^2 + \beta^2 + \gamma^2 - \alpha - \beta - \gamma).
\end{equation}
We note that $a = 0$ implies that $(\alpha^2 + \beta^2 + \gamma^2)f_{\xi} = 0$, which means that either $\alpha = \beta = \gamma = 0$ or $f$ = constant. In both cases, the spatial sections, $t =$ constant, $\psi =$ constant,   are static. Since we are looking for cosmological solutions, \underline{in what follows we will assume $a \neq 0$}. 

From  (\ref{integration of R04}) we obtain 
\begin{equation}
\label{relation between nu and omega in the most general case}
e^{\nu/2} = E f^{b/a}f_{\xi}e^{- \omega/2}.
\end{equation}
Feeding this expression back into (\ref{R11 = R22= R33}) we get
\begin{equation}
\label{relation between f and omega for the most general case}
\left[(b + c)\frac{f_{\xi}}{f}  + a\left(\frac{f_{\xi \xi}}{f_{\xi}} - \frac{\omega_{\xi}}{2}\right)\right]f^{2b/a}f_{\xi}^2 + \epsilon\left[c\left(\frac{f_{\xi}}{f}\right) + \frac{a \omega_{\xi}}{2}\right]\left(\frac{{\cal{E}}}{k E}\right)^2 e^{2 \omega} = 0,
\end{equation}
where 
\begin{equation}
\label{definition of c}
c \equiv \alpha \beta + \alpha \gamma + \beta \gamma.
\end{equation}
Now $R_{0}^{0}$ 
 and $R_{4}^{4}$, depend on the second derivative of $\nu$. Therefore,  after we substitute  (\ref{relation between nu and omega in the most general case}) into them, they become functions of $f_{\xi \xi \xi}$ and $f_{\xi \xi}$, the third and second derivative of $f$. If we isolate $f_{\xi \xi}$ from (\ref{relation between f and omega for the most general case}); calculate the third derivative and substitute into $R_{0}^{0}$ and $R_{4}^{4}$ we find that they vanish identically. 

Consequently, the field equations $R_{AB} = 0$ reduce to two independent equations, namely  (\ref{relation between nu and omega in the most general case}) and (\ref{relation between f and omega for the most general case}) for three unknown metric functions: $\nu(\xi)$, $\omega(\xi)$
 and $f(\xi)$. Thus, fixing one of them we obtain the other two. The interesting point here is that the spacetime metric is completely determined 
by 
the dynamics of the extra dimension. The opposite is also true, namely, that knowing the metric in $4D$ we can reconstruct the geometry in $5D$.

Thus,  in order to obtain specific wave-like solutions one has to complement these equations  with some additional information. The question is how to do that in a way that is physically ``justifiable". The next two sections are devoted to the discussion of this question. 

\section{Solutions  generated from geometrical considerations in $5D$}

In this section we present a number of  solutions to the above equations that follow from the choice of the coordinate/reference system, and the  metric, that are frequently encountered in the literature.   

\subsection{Gaussian normal coordinate system}
A popular choice in the literature is to use the five degrees of coordinate freedom to set $g_{4 \mu} = 0$ and $g_{44} = \epsilon$. This is the so-called ``Gaussian normal coordinate system" based on $\Sigma_{\psi}$.  
In these coordinates we should set $\omega = 0$ in (\ref{relation between f and omega for the most general case}), after which  we obtain a second order differential equation for $f$, 
\begin{equation}
\label{diff equation for f, omega = 0}
a f f_{\xi \xi} + (b + c) f_{\xi}^2 + \epsilon c \left(\frac{{\cal{E}}}{k E}\right)^2 f^{- 2b/a} = 0,
\end{equation}
whose first integral is given by  

\begin{equation}
\label{f for omega = 0}
f_{\xi}^2 = C_{1} f^{- 2(b + c)/a} - \epsilon  \left(\frac{{\cal{E}}}{k E}\right)^2 f^{- 2b/a}, 
\end{equation}
where $C_{1}$ is a constant of integration. 
Consequently, the metric 
\begin{equation}
\label{solution for omega = 0}
dS^2 = E^2\left[C_{1}f^{- 2c/a} - \epsilon\left(\frac{{\cal{E}}}{k E}\right)^2\right]dt^2 - A f^{2 \alpha} dx^2 - B f^{2 \beta}dy^2 - C f^{2 \gamma}dz^2 + \epsilon d\psi^2
\end{equation}
is a solution of the field equations provided the function $f(\xi)$ satisfies (\ref{f for omega = 0}).

\subsection{Synchronous reference system}

The choice $g_{00} = 1$ is usual in cosmology; it corresponds to the so-called synchronous reference system where the time coordinate $t$ is the proper time at each point. 
In order to generate the appropriate solution, let us note that (\ref{integration of R04}) is invariant under the change $\nu \leftrightarrow \omega $. Also, the similarity variable $\xi = {\cal{E}}t + k \psi$ is invariant under the simultaneous change  $t \leftrightarrow \psi$ and $ {\cal{E}} \leftrightarrow k$. Therefore, the line element

\begin{equation}
\label{solution for nu = 0}
dS^2 = dt^2  - A f^{2 \alpha} dx^2 - B f^{2 \beta}dy^2 - C f^{2 \gamma}dz^2 + \epsilon E^2\left[C_{1}f^{- 2c/a} - \epsilon\left(\frac{k}{{\cal{E}} E}\right)^2\right] d\psi^2,
\end{equation}
is also a solution of the field equations $R_{AB} = 0$,  provided  $f(\xi)$ satisfies the equation (\ref{f for omega = 0}) with $ {\cal{E}} \leftrightarrow k$. Namely, 
\begin{equation}
\label{f for nu = 0}
f_{\xi}^2 = C_{1} f^{- 2(b + c)/a} - \epsilon  \left(\frac{k}{{\cal{E}} E}\right)^2 f^{- 2b/a}.
\end{equation}

\subsection{Power-law type solutions}

A simple inspection of the field equations reveals that there are two simple power-law type solutions. They  correspond to the cases where $\nu = \pm \; \omega$.

\subsubsection{Solutions with  $\nu =  \; \omega$}

For $\nu = \omega$  there two different solutions depending on whether $(\epsilon {\cal{E}}^2 + k^2) \neq 0$ or $(\epsilon {\cal{E}}^2 + k^2) = 0$.

\begin{enumerate} 

\item For $(\epsilon {\cal{E}}^2 + k^2) \neq 0$, we substitute $\nu = \omega$ into (\ref{R11 = R22= R33}) and  obtain a simple equation for $f$ whose solution is 
\begin{equation}
\label{f for solution nu = omega}
f(\xi) = \left(c_{1}\xi + c_{2}\right)^{1/a},
\end{equation}
where $c_{1}$ and $c_{2}$ are constants of integration. Feeding back into (\ref{integration of R04}) we find that $e^{\nu} \sim  e^{\omega} \sim f^{- 2 c/a}$.
In summary, the line element 
\begin{equation}
\label{line element for solution nu = omega}
dS^2 = M f^{- 2 c/a} dt^2 - A f^{2 \alpha} dx^2 - B f^{2 \beta}dy^2 - C f^{2 \gamma}dz^2 + \epsilon N f^{- 2 c/a} d\psi^2,
\end{equation}
where $M$ and $N$ are constants, and $f$ satisfies (\ref{f for solution nu = omega}), is a solution of the field equations $R_{AB} = 0$.

\item For $(\epsilon {\cal{E}}^2 + k^2) = 0$, which might happen only if the extra dimension is spacelike $(\epsilon = - 1)$ and ${\cal{E}} = \pm \;k$, equation (\ref{R11 = R22= R33}) is identically satisfied.  Consequently, the line element 
\begin{equation}
\label{solution with epsilon = -1}
dS^2 = E f^{b/a}f_{\xi}dt^2 - A f^{2\alpha}dx^2 - B f^{2 \beta}dy^2 - C f^{2 \gamma}dz^2 - E f^{b/a}f_{\xi}d\psi^2,
\end{equation}
is an exact solution of the field equations $R_{AB} = 0$ for an {\it arbitrary} function $f = f(\xi)$, where  $\xi = {\cal{E}}(t \pm \psi)$, and a general choice of the parameters $\alpha, \beta, \gamma$. In order to avoid misunderstanding, let us emphasize that here $f$ is arbitrary, i.e.,  (\ref{solution with epsilon = -1}) is not necessarily a power-law solution (but we include it here because it belongs to the class of solutions with $\nu = \omega$).
\end{enumerate}

\subsubsection{Solution with  $\nu = - \; \omega$}

If $\omega = - \nu$, then from (\ref{integration of R04}) we obtain a first order differential equation for $f$ whose solution is
\begin{equation}
\label{f for the particular sol}
f(\xi) = \left(C_{2}\xi + C_{3}\right)^{a/(a + b)}, 
\end{equation}
where $C_{2} \equiv [(a + b)/a E]$ and $C_{3}$ is a new constant of integration. For this expression, the field equation (\ref{R11 = R22= R33}) yields 
\begin{equation}
\label{equation for nu, particular solution}
\left(k^2 e^{\nu} - \epsilon {\cal{E}}^2 e^{- \nu}\right)(a + b)\;(C_{2}\xi + C_{3})\;\nu_{\xi} + 2\; c \;C_{2}\left(k^2 e^{\nu} + \epsilon {\cal{E}}^2 e^{- \nu}\right) = 0,
\end{equation}
from which we get
\begin{equation}
\label{Enu for the particular solution}
k^2 e^{\nu} + \epsilon  {\cal{E}}^2 e^{- \nu} = \mbox{constant} \times \left(C_{2} \; \xi + C_{3}\right)^{- 2c/(a + b)}.
\end{equation}
Now, it is not difficult to verify that the line element 
\begin{equation}
\label{metric for the particular solution}
dS^2 = e^{\nu(\xi)}dt^2 - A f^{2\alpha}dx^2 - B f^{2\beta}dy^2  - C f^{2\gamma}dz^2  + \epsilon e^{- \nu(\xi)}d\psi^2,
\end{equation}
with the metric functions $f(\xi)$ and $e^{\nu(\xi)}$, given by (\ref{f for the particular sol}) and (\ref{Enu for the particular solution}), is an exact  solution of the $5D$ field equations $R_{AB} = 0$.
We note that, from (\ref{equation for nu, particular solution}) it follows that the case where $\nu$ = $- \omega = $ constant requires $C_{2} = 0$, i.e., $f = $ constant, in which case the $5D$ manifold is a Minkowski space.

\section{Solutions generated from physical considerations in $4D$}

In this section we derive  a number of  solutions to the $5D$ field  equations that follow from the properties of matter in $4D$.  
In order to make the paper self-consistent, let us restate some concepts that are essential in our discussion. Following the discussion in \cite{Anis Cosm I},  there are different ways of producing, or embedding, a $4D$ spacetime in a given five-dimensional manifold (see e.g., \cite{JPdeLDynamicalFoliationI}). However, the most popular approach is based on three different assumptions. First, that we can use the coordinate frame \cite{Coord. frame}. Second, that our $4D$ spacetime can be recovered by going onto a hypersurface $\Sigma_{\psi}: \psi =  \psi_{0} = $ constant, which is orthogonal to the $5D$ unit vector
\begin{equation}
\label{unit vector n}
{\hat{n}}^{A} = \frac{\delta^{A}_{4}}{\sqrt{\epsilon g_{44}}}, \;\;\;n_{A}n^{A} = \epsilon,
\end{equation}
along the extra dimension. Third, that the physical metric of the spacetime can be identified with  the one induced on $\Sigma_{\psi}$.  

For a line element of the form 
\begin{equation}
\label{General line element in 5D without off-diagonal terms}
dS^2 = g_{\mu\nu}(x^{\rho}, \; \psi)dx^{\mu}dx^{\nu} + \epsilon \Phi^2(x^{\rho}, \; \psi)d\psi^2,
\end{equation}
 the induced metric on hypersurfaces $\Sigma_{\psi}$ is just $g_{\mu\nu}$, i.e., the $4D$ part of the metric in $5D$. The crucial moment is that, although the energy-momentum tensor (EMT) in $5D$ is zero, to an observer  confined to making physical measurements in our ordinary spacetime, and not  aware of the extra dimension,  the spacetime is not empty but contains (effective) matter whose EMT,  ${^{(4)}T}_{\alpha\beta}$, is determined by the Einstein equations in $4D$, namely
\begin{eqnarray}
\label{4D Einstein with T and K}
{^{(4)}G}_{\alpha\beta} = 8 \pi \;{^{(4)}T}_{\alpha\beta} = 
\epsilon\left(K_{\alpha\lambda}K^{\lambda}_{\beta} - K_{\lambda}^{\lambda}K_{\alpha\beta}\right) + \frac{\epsilon}{2} g_{\alpha\beta}\left(K_{\lambda\rho}K^{\lambda\rho} - (K^{\lambda}_{\lambda})^2 \right) - \epsilon E_{\alpha\beta}, 
\end{eqnarray}
where $K_{\mu\nu}$ is the extrinsic curvature 
\begin{equation}
\label{extrinsic curvature}
K_{\alpha\beta} = \frac{1}{2}{\cal{L}}_{\hat{n}}g_{\alpha\beta} = \frac{1}{2\Phi}\frac{\partial{g_{\alpha\beta}}}{\partial \psi};  
\end{equation}
$E_{\mu\nu}$ is the projection of the $5D$ Weyl tensor ${^{(5)}C}_{ABCD}$ orthogonal to ${\hat{n}}^A$, i.e., ``parallel" to spacetime, viz.,
\begin{equation}
\label{Weyl Tensor}
E_{\alpha\beta} = {^{(5)}C}_{\alpha A \beta B}{\hat{n}}^A{\hat{n}}^B 
= - \frac{1}{\Phi}\frac{\partial K_{\alpha\beta}}{\partial \psi} + K_{\alpha\rho}K^{\rho}_{\beta} - \epsilon \frac{\Phi_{\alpha;\beta}}{\Phi},
\end{equation}
and $\Phi_{\alpha} \equiv \partial \Phi/\partial x^{\alpha}$. It is worth mentioning that the effective matter content of the spacetime is the same whether we interpret it in space-time-matter theory, or in a ${\bf Z}_2$ symmetric brane universe \cite{physical motivation}. 

\subsection{Properties of the effective energy-momentum tensor}

For the case under consideration, the spacetime metric induced on $\Sigma_{\psi}$ is given by
\begin{equation}
\label{spacetime metric}
ds^2 \equiv dS^2_{|\Sigma_{\psi}} = g_{\mu\nu}dx^{\mu}dx^{\nu} = E^2 f^{2b/a}f_{\xi}^2 e^{- \omega(\xi)}dt^2 - A f^{2\alpha}dx^2 - B f^{2\beta}dy^2 - C f^{2\gamma}dz^2.
\end{equation}
The effective EMT is\footnote{To simplify the notation, in what follows we suppress the index $^{(4)}$ in $^{(4)}T_{\alpha\beta}$}

\begin{eqnarray}
\label{general T00, T11, T22, T33}
T_{0}^{0} &=& \frac{c {\cal{E}}^2 e^{\omega}}{E^2 f^{2(a + b)/a}},\;\;\;\;c \neq 0, \nonumber \\
T_{1}^{1} &=& T_{0}^{0}\left[\frac{\beta + \gamma - \alpha}{a} + \frac{\omega_{\xi} (\beta + \gamma)f}{2 c f_{\xi}}\right],\nonumber \\
T_{2}^{2} &=& T_{0}^{0}\left[\frac{\alpha + \gamma - \beta}{a} + \frac{\omega_{\xi} (\alpha + \gamma)f}{2 c f_{\xi}}\right],\nonumber \\
T_{3}^{3} &=& T_{0}^{0}\left[\frac{\alpha + \beta - \gamma}{a} + \frac{\omega_{\xi} (\alpha + \beta)f}{2 c f_{\xi}}\right].
\end{eqnarray}
Let us notice that, for a general choice of parameters, the components of the EMT satisfy the following algebraic expressions 

\begin{eqnarray}
(\beta - \gamma)(T_{0}^{0} + T_{1}^{1}) + (\beta +\gamma)(T_{2}^{2} - T_{3}^{3}) &=& 0, \nonumber \\
(\alpha - \gamma)(T_{0}^{0} + T_{2}^{2}) + (\alpha +\gamma)(T_{1}^{1} - T_{3}^{3}) &=& 0, \nonumber \\
(\alpha - \beta)(T_{0}^{0} + T_{3}^{3}) + (\alpha + \beta)(T_{1}^{1} - T_{2}^{2}) &=& 0.
\end{eqnarray}
They can be interpreted as ``equations of state" for the effective density $T_{0}^{0}$ and the stresses $T_{i}^{i}$, with $i = 1, 2, 3$. One can use them to obtain  a simple relationship between the stresses, viz.,
\begin{equation}
\label{relation between spatial T's}
(\beta - \gamma)T_{1}^{1} + (\gamma - \alpha)T_{2}^{2} + (\alpha - \beta) T_{3}^{3} = 0.
\end{equation}
In the case of axial symmetry, say along the $x$-direction ($\beta = \gamma$),   they reduce to 
\begin{equation}
\label{equation of state for beta = gamma}
T_{2}^{2} = T_{3}^{3}, \;\;\;\;\mbox{and}\;\;\;\;   T_{0}^{0} = - \frac{\alpha + \beta}{\alpha - \beta}T_{1}^{1} + \frac{2 \beta}{\alpha - \beta}T_{2}^{2}, \;\;\;\;\alpha \neq \beta = \gamma.
\end{equation}
In the case of isotropic expansion $(\alpha = \beta = \gamma)$ the effective EMT behaves like a perfect fluid 
\begin{equation}
\label{isotropic expansion}
T_{1}^{1} = T_{2}^{2} = T_{3}^{3} = n T_{0}^{0}, \;\;\;\;\mbox{with}\;\;\; n \equiv \frac{1}{3}\left(1 + \frac{\omega_{\xi}f}{\alpha f_{\xi}}\right),
\end{equation}
which for $n = $ constant is nothing but the barotropic equation of state commonly used in cosmological problems. 

\medskip

In general, we can write

\medskip

\begin{equation}
\label{general T00}
T_{1}^{1} = n_{x}T_{0}^{0}, \;\;\;\;\;T_{2}^{2} = n_{y}T_{0}^{0}, \;\;\;\;\;T_{3}^{3} = n_{z}T_{0}^{0}, 
\end{equation}
where 
\begin{enumerate}
 \item For $\beta \neq - \gamma$

\begin{equation}
\label{definition of the n's: beta neq - gamma}
n_{y} \equiv \frac{n_{x}(\alpha + \gamma) + (\alpha - \beta)}{\beta + \gamma}, \;\;\;\;\;n_{z} \equiv \frac{n_{x}(\alpha + \beta) + (\alpha - \gamma)}{\beta + \gamma}, \;\;\;\;\;n_{x} =  \left[\frac{\beta + \gamma - \alpha}{a} + \frac{\omega_{\xi} (\beta + \gamma)f}{2 c f_{\xi}}\right].
\end{equation}

\item For $\beta = - \gamma$ and $\alpha \neq - \beta$

\begin{equation}
\label{definition of the n's : beta = - gamma  and alpha neq - beta}
n_{x} = - 1, \;\;\;n_{y} = \frac{n_{z}(\alpha - \beta) - 2\beta}{\alpha + \beta}, \;\;\;\;n_{z} = \frac{\alpha + 2 \beta}{\alpha} - \frac{(\alpha + \beta)\omega_{\xi} f}{2 \beta^2 f}.
\end{equation}

\item For $\beta = - \gamma$ and $\alpha = - \beta$ $(\alpha = \gamma = - \beta)$
\begin{equation}
\label{definition of n's: beta = - gamma and alpha = - beta}
n_{x} = - 1, \;\;\;\;n_{z} = - 1, \;\;\;\;n_{y} =  3 - \frac{\omega_{\xi} f}{\alpha f_{\xi}}.
\end{equation}

\end{enumerate}
Clearly, one can use the above expressions to obtain the parameters $(\alpha,\; \beta,\; \gamma)$ in terms of $(n_{x},\; n_{y},\; n_{z})$.  It should be emphasized that, in the wave-like cosmologies under study,  the effective matter in $4D$ behaves like a perfect fluid $(n_{x} = n_{y} = n_{z})$ only for isotropic expansion. This is different from other models which do allow perfect fluid  and  anisotropic expansion (see,  e.g., \cite{Suresh} and references therein, also the solution with stiff equation of state given in section $5.1.2$ of our previous work \cite{Anis Cosm I}).

\subsection{Solutions for constant $n_{x}$, $n_{y}$ and $n_{z}$ }

Let us assume that the ratio $n_{i} = T_{i}^{i}/ T_{0}^{0}$ is constant in every direction, which constitutes an extension  to anisotropic models of the assumption of barotropic expansion used in FRW cosmologies. 
Without loss of generality, for the sake of argument let us assume that  $\beta \neq - \; \gamma$. In this case from (\ref{definition of the n's: beta neq - gamma}) it follows that
\begin{equation}
\label{solution with constant equation of state}
e^{\omega} = D f^{2 \eta/a}, \;\;\;\;\mbox{where}\;\;\;\eta \equiv \frac{c \left[(n_{x} - 1)(\beta + \gamma) + (n_{x} + 1)\alpha\right]}{(\beta + \gamma)}, \;\;\;\beta \neq  - \gamma
\end{equation}
and $D$ is a constant of integration. Substituting this into (\ref{relation between f and omega for the most general case}) we get an equation for $f$, viz., 

\begin{equation}
\label{diff equation for f for constant n's}
a f f_{\xi \xi} + (b + c - \eta) f_{\xi}^2 + \epsilon (c + \eta) \left(\frac{{\cal{E}}D}{k E}\right)^2 f^{2(2\eta - b)/a} = 0,
\end{equation}
whose first integral is given by  
\begin{equation}
\label{f for constant n's}
f_{\xi}^2 = C_{1} f^{- 2(b + c - \eta)/a} - \epsilon  \left(\frac{{\cal{E}}D}{k E}\right)^2 f^{2(2\eta - b)/a}, 
\end{equation}
where $C_{1}$ is a constant of integration. Consequently, the metric 
\begin{equation}
\label{solution for constant n's}
dS^2 = \left[{\bar{C}}_{1}f^{- 2c/a} - \epsilon D\left(\frac{{\cal{E}}}{k} \right)^2 f^{2\eta/a}\right]dt^2 - A f^{2 \alpha} dx^2 - B f^{2 \beta}dy^2 - C f^{2 \gamma}dz^2 + \epsilon D f^{2 \eta/a}d\psi^2,
\end{equation}
with ${\bar{C}}_{1} = E^2 C_{1}/D$, is a solution of the $5D$ field equations $R_{AB} = 0$ provided the function $f(\xi)$ satisfies (\ref{f for constant n's}), regardless the signature of the extra dimension and {\it arbitrary} parameters $\alpha$, $\beta$ and $\gamma$.  It is clear that solutions with   $\beta = - \gamma$, as well as the axially symmetric ones with $\beta = \gamma$, can be obtained in a similar way.

It should be  noted that the \underline{solutions in Gaussian normal coordinates (\ref{solution for omega = 0}) are particular cases of (\ref{solution for constant n's}) with $\eta = 0$}. Also, as mentioned above, the parameters can be expressed in terms of the barotropic coefficients. In the present case they are 

\begin{equation}
\label{the parameters in terms of the n's}
\beta = \alpha\; \frac{( 1 + n_{x} - n_{y} + n_{z})}{(1 - n_{x} + n_{y} + n_{z})}, \;\;\;\;\gamma = \alpha\; \frac{( 1 + n_{x} + n_{y} - n_{z})}{(1 - n_{x} + n_{y} + n_{z})}.
\end{equation}
Certainly, without loss of generality one  can add any additional condition on these parameters. 

\subsection{Radiation-like solutions}

  It is well known that in the case of radiation (i.e., photons with zero rest mass) as well as for ultra-relativistic matter (i.e., particles with finite rest masses moving close to the speed of light) the trace of the EMT, say $T$, vanishes identically. A simple calculation from (\ref{general T00, T11, T22, T33}) gives
\begin{equation}
\label{Trace of the EMT}
T = T_{0}^{0} +  T_{1}^{1} +  T_{2}^{2} +  T_{3}^{3} = T_{0}^{0}\left(2 + \frac{a\; \omega_{\xi} f}{c f_{\xi}}\right). 
\end{equation}
Thus, for $T = 0$ (and $T_{0}^{0} \neq 0$) the second term in (\ref{relation between f and omega for the most general case}) vanishes. Consequently, \underline{radiation-like solutions}  have $e^{\omega} \sim f^{- 2c/a}$ and \underline{are given by the power-law line element (\ref{line element for solution nu = omega})}.

\subsection{Vacuum solutions in $4D$}

From (\ref{general T00, T11, T22, T33}) it follows that $T_{\mu\nu} = 0$ requires  $c = 0$ and $\omega_{\xi} = 0$. Thus,  \underline{vacuum solutions are particular cases} (with $c = 0$) \underline{of those in Gaussian coordinates (\ref{solution for omega = 0})}. Then, from (\ref{f for omega = 0}) it follows that  $f(\xi) \sim \xi^{a/(a + b)}$. Therefore, the $5D$ line element 

\begin{equation}
\label{vacuum solutions}
dS^2 = dt^2 - A \xi^{2 p_{1}}dx^2 - B \xi^{2 p_{2}}dy^2 - C \xi^{2 p_{3}}dz^2 + \epsilon d\psi^2,
\end{equation}
with
\begin{equation}
\label{p's for the vacuum solution}
p_{1} = \frac{\alpha \; a}{a + b}, \;\;\;\;p_{2} = \frac{\beta \;  a}{a + b}, \;\;\;\;p_{3} = \frac{\gamma \; a}{a + b},
\end{equation}
which satisfy $p_{1} + p_{2} + p_{3} = 1$ and $p_{1}^2 + p_{2}^2 + p_{3}^2 = 1$, may be interpreted as an embedding for a $4D$ Kasner spacetime. Clearly, vacuum solutions can also be obtained from those in the synchronous reference system (\ref{solution for nu = 0}) by setting  $c = 0$.

\section{Other interpretations in $4D$}

In the case where  the extra dimension is spacelike, the solutions to the $5D$ field equations are invariant under the transformation $(x, y, z) \leftrightarrow \psi$. However, the physics in $4D$  crucially depends on how the coordinates of our ordinary $3D$ space are identified.  In fact, after such a transformation (i) the spacetime slices $\Sigma_{\psi}$ are non-flat, and (ii) the effective four-dimensional EMT is traceless.

As an illustrative simple example, let us consider the $5D$ metric
\begin{equation}
\label{line element for solution nu = omega, after z-psi transformation}
dS^2 = M F^{- 2 c/a} dt^2 - A F^{2 \alpha} dx^2 - B F^{2 \beta}dy^2 - N F^{- 2 c/a} dz ^2  + \epsilon C F^{2 \gamma}d\psi^2,
\end{equation}
with  $F = (c_{1}\eta + c_{2})^{1/a}$ and $\eta = {\cal{E}}t + k z$, which is a solution of the $5D$ field equations $R_{AB} = 0$
 obtained from the power-law solution (\ref{line element for solution nu = omega}) after a $z \leftrightarrow \psi$ transformation. Here the metric functions are independent of the extra coordinate $\psi$ and, consequently, the effective EMT is traceless.  Therefore,   the  $4D$ metric induced on $\Sigma_{\psi}$ can be interpreted either as anisotropic  \underline{vacuum} solutions in the Randall-Sundrum (RS2) braneworld scenario \cite{Anis Cosm I},  or as  inhomogeneous \underline{radiation-like}  solutions in conventional $4D$ general relativity. Certainly, the same is true for all the solutions discussed here.    

The above discussion highlights the fact that  much work is still needed in order to have a clear understanding of $4D$ physical models as Lorentzian hypersurfaces in pseudo-Riemannian $5D$ spaces. 

\section{Summary}

According to Campbell's theorem any solution of the Einstein equations in $4D$, with an arbitrary energy-momentum tensor, can be locally embedded in a solution of the vacuum Einstein field equations in $5D$. In this paper we have considered a five-dimensional Riemannian manifold whose line element is diagonal and independent of coordinates for ordinary $3D$ space $(x, y, z)$, which is given by (\ref{General metric in 5D }). This line element is quite general  in the sense that, on every hypersurface $\Sigma_{\psi}: \psi = \psi_{0}$,  it  can be used or interpreted  as a $5D$  embedding  for spatially flat FRW models,  as well as for Bianchi type-I cosmologies.

In order to solve the field equations $R_{AB} = 0$ we have assumed that the $5D$ manifold  presents self-similar symmetry. In the traditional interpretation of Sedov, Taub and Zeldovich \cite{Sedov}, this means that all dimensionless quantities in the theory can be expressed as functions only of a single similarity variable. This assumption, and the field equations,  determine the possible shape of the similarity variable.  These are  (i) $\xi = t/\psi$; (ii) $\xi = e^{q t}/e^{q \psi}$, and (iii) $\xi = {\cal{E}} t + k \psi$, where $q, {\cal{E}}$ and $k$ are some constants with the appropriate units.
They correspond to three distinct families of self-similar solutions, each of them being  parameterized by an arbitrary function of the similarity variable and three arbitrary parameters $\alpha, \beta, \gamma$.

\begin{enumerate}
  
  \item For $\xi = t/\psi$ the $5D$ manifold possesses homothetic symmetry and the general  solution of the field equations $R_{AB} = 0$ is given by \cite{Anis Cosm I}  
\begin{equation}
\label{in summary}
dS^2 = E f^{(b/a)}f_{\xi}dt^2 - A f^{2 \alpha}dx^2 - B f^{2 \beta}dy^2 - C f^{2 \gamma}dz^2 - E \xi^2 f^{(b/a)}f_{\xi}d\psi^2. 
\end{equation}

\item For $\xi = \left(e^t/e^\psi\right)^q$ the $5D$ manifold possesses conformal  symmetry and the general  solution of the field equations $R_{AB} = 0$ is  \cite{Anis Cosm I}  
\begin{equation}
\label{general solution for l = 1}
dS^2 = {E} \xi f^{(b/a)}f_{\xi}dt^2 - A f^{2 \alpha}dx^2 - B f^{2 \beta}dy^2 - C f^{2 \gamma}dz^2 - E \xi f^{(b/a)}f_{\xi}d\psi^2.
\end{equation}
\item For $\xi = {\cal{E}} t + k \psi$, in this work we have seen that the general solution can be written as 
\begin{equation}
\label{general solution for the wave-like case}
dS^2 = {E}^2 f^{(2 b/a)}f_{\xi}^2 e^{- \omega} dt^2 - A f^{2 \alpha}dx^2 - B f^{2 \beta}dy^2 - C f^{2 \gamma}dz^2  + \epsilon e^{\omega}d\psi^2,
\end{equation}
where the functions $f$ and $\omega $ are related by (\ref{relation between f and omega for the most general case}). 
This case is mathematically more complicated than the other two because, after choosing some specific  function, say  $f$  for the sake of argument, one still has to integrate (\ref{relation between f and omega for the most general case}) in order   to concretize   the solution. 

\end{enumerate} 

For the wave-like models discussed here, as well as for the ones discussed in our recent work \cite{Anis Cosm I}, the arbitrary function needed to specify the solution can be determined either by the choice of the reference/coordinate system, as it is illustrated in section $3$, or by imposing certain conditions on the effective EMT in $4D$, as we did in section $4$. In all cases the parameters $\alpha, \beta$ and $\gamma$ are related to the properties of the effective $4D$ matter. They are not independent and, therefore,  without loss of generality, one can impose any algebraic condition on them. This is illustrated by (\ref{the parameters in terms of the n's}), as well as by the solutions discussed in section $5.1$ of  \cite{Anis Cosm I}. 

According to Campbell's theorem the connection to $4D$ is deduced after choosing an embedding\footnote{From a general-relativistic  viewpoint one could argue that, instead of deriving the properties of the EMT from an embedding  in $5D$, a more `physical' approach would be to find an embedding for a $4D$ spacetime with a specified physical EMT. However, this seems to  face fundamental problems \cite{Anis Cosm I}.} . In order to keep the spacetime signature $(+,   -,   -,   -)$ in (\ref{in summary})-(\ref{general solution for the wave-like case}) the constants $A- E$ ought to be positive, otherwise they are arbitrary. Thus, the extra dimension must be spacelike for the homothetic and conformal solutions and, in general,  it is undefined $(\epsilon = \pm 1)$ for wave-like solutions. Although an exception is provided by solution (\ref{solution with epsilon = -1}) which requires  $\epsilon = - 1$.

For wave-like solutions the effective matter in $4D$ cannot be interpreted as perfect fluid, except in the isotropic limit corresponding to $\alpha = \beta = \gamma$. This is quite different from the homothetic solutions (\ref{in summary}) which do allow such interpretation (section $5.1$ in \cite{Anis Cosm I}). 
Despite these differences,  the anisotropic $5D$ cosmologies  (\ref{in summary})-(\ref{general solution for the wave-like case}) share in common the property that, although they are Ricci-flat $(R_{AB} = 0)$, they are not Riemann-flat $(R_{ABCD} = 0)$, except in the trivial case where $a = 0$. The distinction is important because certain solutions of $5D$ relativity with high degrees of symmetry may have $R_{ABCD} = 0$ and be flat in $5D$, while possessing curved subspaces in $4D$. This is the case of $5D$ cosmologies with spherical symmetry in ordinary $3D$ space \cite{JACP}, which include the standard $5D$ cosmologies \cite{standard}. 

Our solutions may be applied to the era after the big-bang,  where the anisotropy could have played  a significant role,  and the universe could not have been well described by FRW isotropic models. Besides, the solutions allow different equations of state, among them a radiation-like equation of state, with $T = 0$, which is typical of radiation and/or ultra-relativistic matter.

\end{document}